\documentclass[twoside]{dis09}
\usepackage[latin1]{inputenc}
\usepackage[dvips]{graphicx,epsfig,color}
\usepackage{wrapfig,rotating}
\usepackage{amssymb,amsmath,array}

\def\eq#1{{Eq.~(\ref{#1})}}

\begin{document}
\title{A Global Analysis of DIS Data at Small-x with Running Coupling
  BK Evolution}

\author{\underline{Javier L. Albacete}$^{1,2}$, N\'estor Armesto$^3$,
  Jos\'e Guilherme Milhano$^4$ and Carlos A. Salgado$^3$
\vspace{.3cm}\\
1- ECT*, Strada delle Tabarelle 286, I-38050
  Villazzano (TN), Italy
\vspace{.1cm}\\
2- Institute de Physique Th\`eorique, CEA Saclay (URA 2306),
F-91191,\\
Gif-sur-Yvette, France 
\vspace{.1cm}\\
3-Departamento de F\'{\i}sica de Part\'{\i}culas and
IGFAE,
Universidade de Santiago de Compostela,\\ E-15706 Santiago de Compostela, Spain
\vspace{.1cm}\\ 
4- CENTRA, Instituto Superior T\'ecnico (IST),
Av. Rovisco Pais, P-1049-001 Lisboa, Portugal}

\maketitle

\begin{abstract}
  We present a global fit to the structure function $F_2$ measured in
  lepton-proton experiments at small values of Bjorken-$x$, $x\le
  0.01$, for all experimentally available values of $Q^2$, $0.045 \
  {\rm GeV}^2\le Q^2 \le 800 \ {\rm GeV}^2$, using the Balitsky-Kovchegov
  equation including running coupling corrections. Using our fits to
  F$_2$, we reproduce available data for F$_L$ and perform
  predictions, parameter-free and completely driven by small-x
  evolution, to the kinematic range relevant for the LHeC.
\end{abstract}

Deep Inelastic lepton-hadron Scattering (DIS) experiments constitute
one of the main sources of information about the partonic structure of
hadrons, as well as a testing ground of pQCD techniques. The DIS
program carried out in HERA has explored an unprecedented kinematic
regime in the DIS variables $Q^2$ and Bjorken-$x$. At small values of
$x$, the emission of additional soft gluons in the hadron wavefunction
is enhanced by factors of $\alpha_s\,\ln 1/x\sim \mathcal{O}(1)$,
resulting in a fast growth of the gluon density of the hadron with
decreasing Bjorken-$x$. Such growth has been observed in data and, at
small enough values of $x$, or, equivalently, for large gluon
densities, it is expected to be tamed by repulsive gluon-gluon
self-interactions. The latter mechanism is known as {\it saturation}
of the gluon distribution. It is non-linear phenomenon, closely related
to unitarity of the theory, which is not included in the more standard
linear approaches to DIS based on the DGLAP or BFKL pQCD evolution
equations. Here we present an analysis of available DIS data at
small-$x$ based on the non-linear Balitsky-Kovchegov (BK) equation
including the recently calculated running coupling corrections to the
evolution kernel
\cite{Kovchegov:2006vj,Balitsky:2006wa,Gardi:2006rp,Albacete:2007yr}.

The starting point of our analysis is the dipole model of DIS, which
stems from the $k_t$-factorization theorems in the high-energy
limit. It states that, at small-$x$, the virtual photon-proton cross
section can be written as a convolution of the light-cone wavefunction
for a virtual photon to split into a quark-antiquark dipole,
$|\Psi_{T,L}|^2$, times the cross section for dipole-proton
interaction. For a hadron homogeneous in the transverse plane, one has:
\begin{equation}
  \sigma_{T,L}^{\gamma^*p}(x,Q^2)=\sigma_0\,\int_0^1 dz\,d{\bf r}\,\vert
  \Psi_{T,L}(z,Q^2,{\bf r})\vert^2\,
  {\cal N}(r,x)\,,
\label{dm1}
\end{equation}
where the subscripts T, L stand for the polarization of the virtual
photon (transverse or longitudinal), $r$ is the dipole transverse size
and $Q^2$ the photon virtuality. Explicit expressions for $\Psi_{T,L}$
in the case of any number of active flavors can be found in
\cite{Golec-Biernat:1998js}. In our analysis we consider only the
three flavor case, i.e, with no charm quark. All the
information about the strong interactions, and also all the $x$
dependence of the absorption cross section, is encoded in the
imaginary part of the dipole scattering amplitude, ${\mathcal N}(r,x)$
(related to the dipole cross section through the optical theorem). The
normalization $\sigma_0$ in \eq{dm1} originates from integration over
the transverse plane, trivial in the case of a {\it cylindrical}
proton considered here, and will be one of the free parameters in the
fit. The inclusive DIS structure function is then given by
\begin{equation}
F_2(x,Q^2)=\frac{Q^2}{4\,\pi^2\alpha_{em}}\left(\sigma_T+\sigma_L\right)\,,
\label{f2}
\end{equation}
whereas the longitudinal structure function $F_L(x,Q^2)$ is obtained
by just considering the longitudinal component.

General arguments in QCD indicate that the dipole scattering amplitude
is small for small dipole sizes $r$ ({\it color transparency}),
whereas large dipoles suffer a large absorption. Phenomenological
works in the context of saturation parametrize the transition between
these two regimes, {\it dilute} and {\it saturated}, in terms of a
$x$-dependent saturation scale $Q_s^2(x)$, such that
$\mathcal{N}(r,x)\sim r^2$ for $r\ll 1/Q_s(x)$, whereas
$\mathcal{N}(r,x)\sim 1$ for $r\gg1/Q_s$. The details of the
parametrization depend on the physical input considered in each
case. The leading $x$-dependence of the structure functions is thus
encoded in that of the saturation scale, commonly assumed to have a
power-law behavior: $Q_s^2(x)\sim(x_0/x)^{\lambda}$ GeV$^2$, with the
{\it evolution speed} $\lambda$ being a free parameter fitted to
experimental data. The results from different groups, albeit using
slightly different parametrizations of the dipole amplitude, report
$\lambda\sim 0.2\div 0.3$.

In this work, rather than resorting to phenomenological models, we
calculate the $x$-dependence of the dipole amplitude in terms of the
pQCD non-linear BK evolution equation including the recently
calculated running coupling corrections. Importantly, the original,
leading-log (LL) in $\alpha_s\ln1/x$ BK equation yields an evolution
speed $\lambda\sim 4.8\, \alpha_s$, with $\alpha_s$ the bare coupling
\cite{Albacete:2004gw}. Such value is too large to attempt a good
description of data. Importantly, the running coupling corrections to
the LL kernel obtained through resummation of $\alpha_s\,N_f$ terms
bring the evolution speed down to values compatible with those
extracted from data, among other interesting dynamical effects
\cite{Albacete:2007yr}. The BK equation reads
\begin{eqnarray}
  \frac{\partial{\cal{N}}(r,Y)}{\partial\,\ln x_0/x}&=& \int d{\bf r_1}\,
  K({\bf r},{\bf r_1},{\bf r_2})\nonumber \\
  &\times&
  \left[{\cal N}(r_1,x)+{\cal N}(r_2,x)-{\cal N}(r,x)-
    {\cal N}(r_1,x)\,{\cal N}(r_2,x)\right]\,,
\label{bklo}
\end{eqnarray}
where $\bf{r_2}=\bf{r}-\bf{r_1}$ and $x_0$ is the starting point for
the evolution. In our case $x_0=10^{-2}$ is the largest experimental
value of $x$ included in the fit. The precise shape of the running
coupling kernel $K$ depends on the scheme chosen to single out the
ultra-violet divergent contributions from the finite ones that
originate after the resummation of quark loops. It was found in
\cite{Albacete:2007yr} that the prescription suggested by Balitsky in
\cite{Balitsky:2006wa} minimizes the role of the subleading
contributions, and is therefore better suited for phenomenological
applications. The corresponding kernel is
\begin{equation}
  K({\bf r},{\bf r_1},{\bf r_2})=\frac{N_c\,\alpha_s(r^2)}{2\pi^2}
  \left[\frac{r^2}{r_1^2\,r_2^2}+
    \frac{1}{r_1^2}\left(\frac{\alpha_s(r_1^2)}{\alpha_s(r_2^2)}-1\right)+
    \frac{1}{r_2^2}\left(\frac{\alpha_s(r_2^2)}{\alpha_s(r_1^2)}-1\right)
  \right]\,.
\label{kbal}
\end{equation}
\begin{table}[h]
\begin{center}
\begin{tabular}{c|c|c|c|c|c}
  Initial condition & $\sigma_0$ (mb) & $Q_{s0}^2$ (GeV$^2$) &
  $C^2$ &$\gamma$ & $\chi^2/{\rm d.o.f.}$ \\
  \hline
  GBW & 31.59 & 0.24 & 5.3 & 1 (fixed) & 916.3/844=1.086 \\
  \hline
  MV & 32.77 & 0.15 & 6.5 & 1.13 & 906.0/843=1.075\\
   \hline
\end{tabular}
\label{restab}
\caption{Values of the fitting parameters from the fit to available $F_2(x,Q^2)$
  data with $x\leq10^{-2}$ and for all
  available values of $Q^2$, $0.045 \ {\rm GeV}^2\le Q^2 \le 800 \
  {\rm GeV}^2$.}
\end{center}
\end{table}
\hskip -0.2cm \noindent Details about the numerical method used to solve the BK equation can
be found in \cite{Albacete:2009fh}. We evaluate the coupling according
to the standard one-loop QCD expression. In order to regularize it in
the infrared, we freeze it to a constant value for $r>r_{fr}$, with
$\alpha_s(r_{fr}^2)=0.7$. Thus
\begin{equation}
\alpha_s(r^2)=\frac{12\pi}{\left(11N_c-2N_f\right)\,\ln\left
    (\frac{4\,C^2}{r^2 \Lambda_{QCD}^2}\right)}\,\,\,\text{for}\,\,r<r_{fr},
\label{alpha}
\end{equation}
and $\alpha_s(r^2)=0.7$ for $r>r_{fr}$. The parameter $C$ in
\eq{alpha} is a free parameter in the fit. We take
$\Lambda_{QCD}=0.241$ GeV.
\begin{figure}
\begin{center}
\includegraphics[width=8.5 cm]{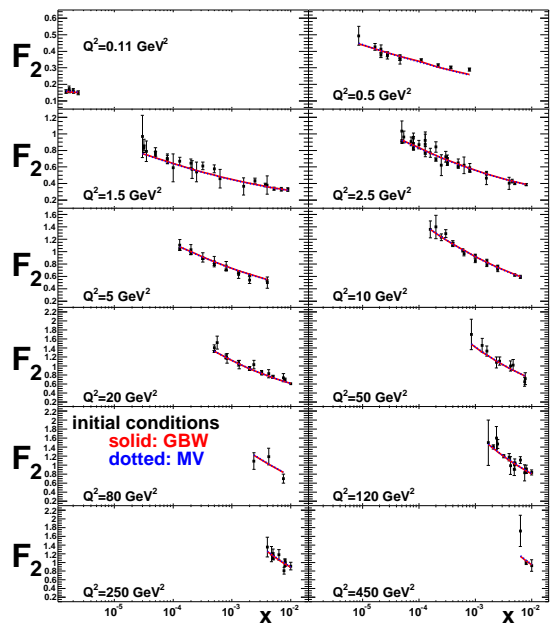}
\end{center}
\caption{Comparison between a selection of experimental data
  and the results from the fit for
  $F_2(x,Q^2)$. Solid red lines correspond to GBW i.c., and dotted
  blue ones to MV i.c.}
\label{figf2}
\end{figure}
A final ingredient in our analyses is the initial condition for the
evolution, i.e. the specific form of the dipole amplitude at
$x=10^{-2}$. We considered two different initial
conditions, {\it GBW}:
\begin{equation}
 {\cal N}^{GBW}(r,x=10^{-2})=
1-\exp{\left[-\left(\frac{r^2\,Q_{s\,0}^2}{4}\right)^{\gamma\,}\right]}\,,
\label{gbw}
\end{equation} 
and {\it MV}:
\begin{equation} 
{\cal
  N}^{MV}(r,x=10^{-2})=1-\exp{\left[-\left(\frac{r^2Q_{s\,0}^{2}}{4}\right)^{\gamma}
    \ln{\left(\frac{1}{r\,\Lambda_{QCD}}+e\right)}\right]}\, ,
\label{mv}
\end{equation} where $Q_{s\,0}^2$ is the initial saturation scale and $\gamma$
is another free parameter which controls the small $r$ behavior of
the dipole amplitude. For the fit with {\it GBW} i.c. $\gamma$ is kept
fixed to 1. 

\begin{figure}[ht]
\begin{center}
\hskip -0.2cm \includegraphics[width=0.5\columnwidth]{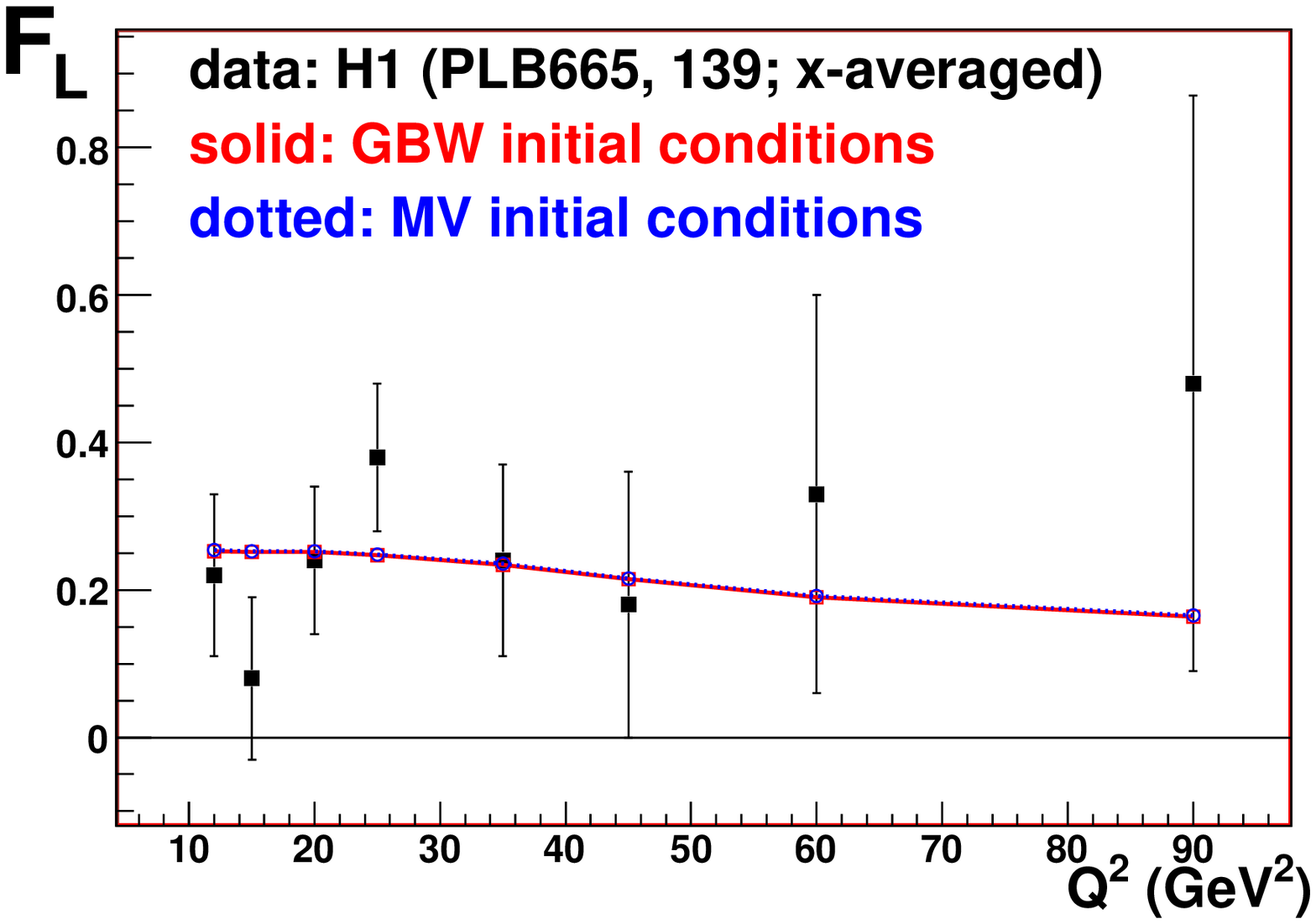}
\includegraphics[width=0.5\columnwidth]{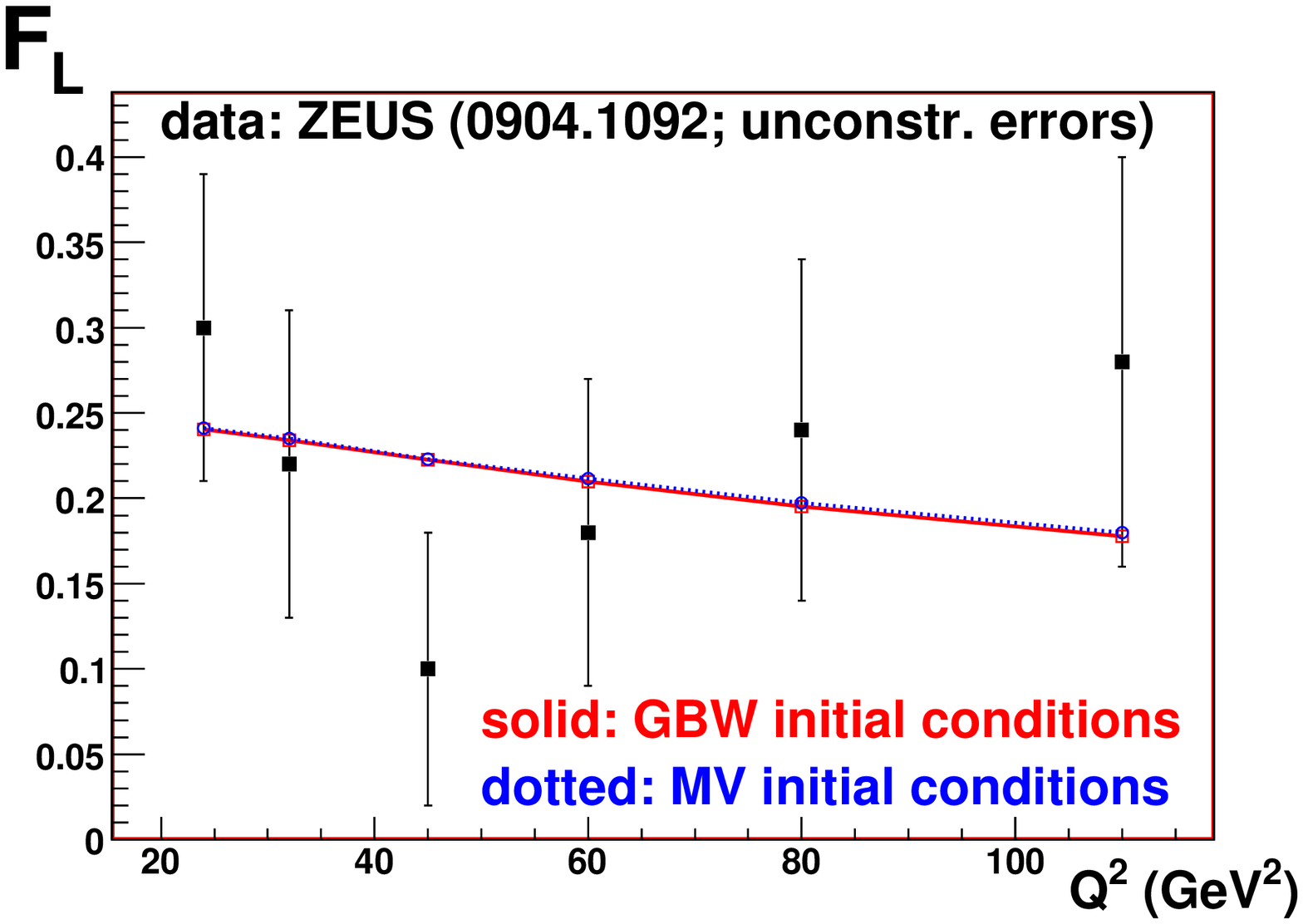}
\end{center}
\caption{Comparison between experimental data from the H1 (left) and
  ZEUS (right) and the predictions of our model for $F_L(x,Q^2)$. The
  theoretical results have been computed at the same $\langle
  x\rangle$ as the experimental ones.}
\label{figfl}
\end{figure}

All in all, there are four (three when $\gamma$ is fixed) free parameters
in the fit: The initial saturation scale $Q_{s0}$, the overall
normalization $\sigma_0$, the infrared parameter $C$ and the anomalous
dimension $\gamma$.  With this setup we obtain an excellent fit to
available experimental data for the inclusive DIS structure
function $F_2(x,Q^2)$ with $x<10^{-2}$ and all existing values for
$Q^2$, $0.045\leq Q^2 \leq 800$ GeV$^2$, which adds up to 847 data
points.  In order to smoothly go to photoproduction, we have used the
following redefinition of the Bjorken variable:
$\tilde{x}=x\,(1+4m_f^2/Q^2)$, with $m_f=0.14$ GeV for the three light
flavors considered in this work. The experimental data included in the
fit has been obtained by the H1 and ZEUS Collaborations (HERA) and by
the E665 (Fermilab) and NMC (CERN) Collaborations\footnote{A detailed
  list of references to the publications where data is presented can
  be found in \cite{Albacete:2009fh}}. The parameters from the fit are
presented in Table 1. We have checked that the quality of the fit and
the values of the parameters are stable under the restriction of the
data range to the region $Q^2< 50$ GeV$^2$ (which leaves 703 data
points for the fit). A comparison between a selection of experimental
data and the results from the fit for $F_2(x,Q^2)$ is shown in
Fig. \ref{figf2}. The two initial conditions {\it GBW} and {\it MV}
offer equally good fits to data. With the parameters values yielded by
the fit, we obtain a good description of the longitudinal structure
function, as shown in Fig. \ref{figfl}.

\begin{figure}
\hskip -1cm \includegraphics[width=0.55\columnwidth]{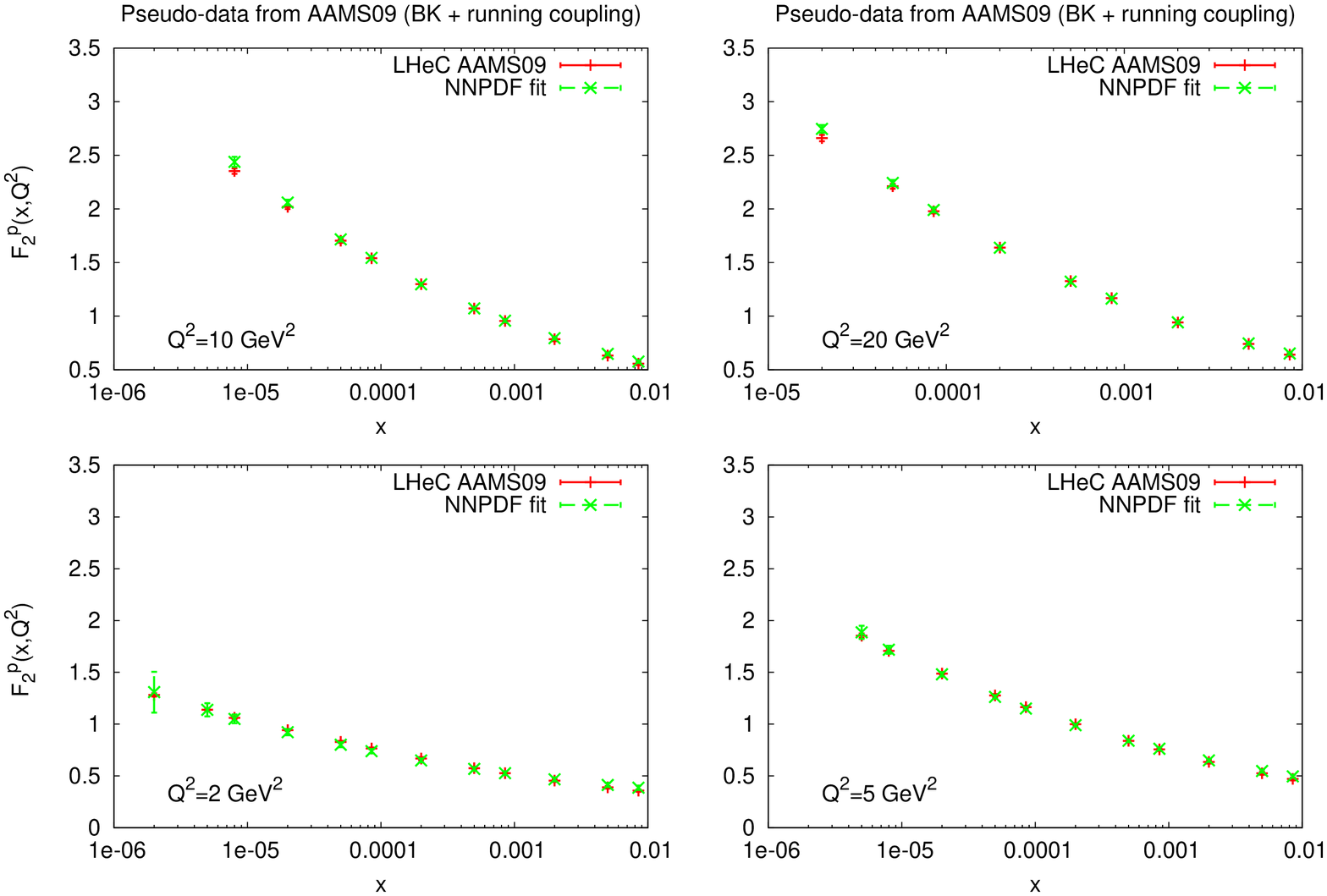}
\includegraphics[width=0.55\columnwidth]{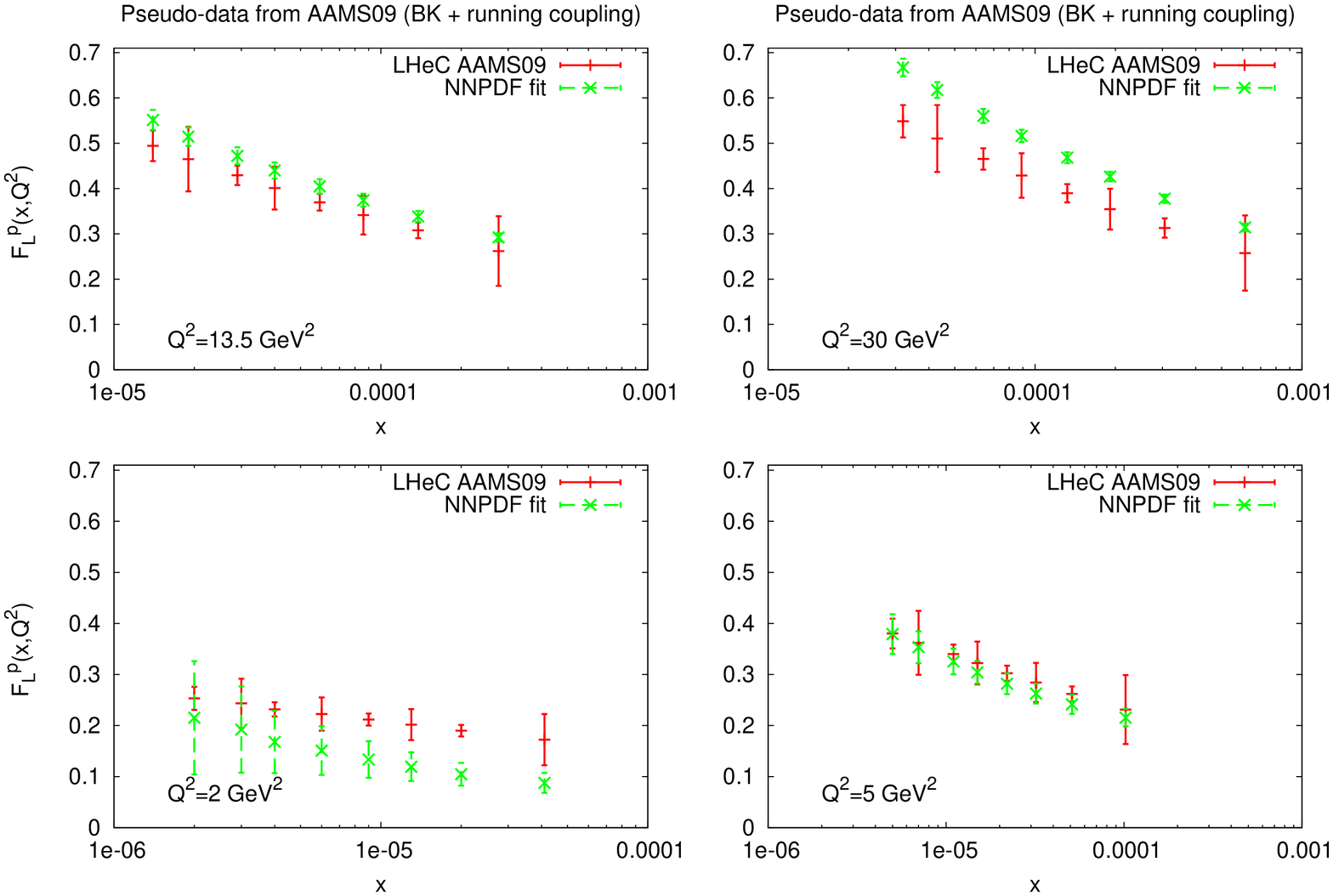}
\caption{NNPDF1.2 DGLAP fits (green stars) to pseudodata (red bars) for
  proton F$_2$ (left plot) and F$_L$ (right plot) generated by
  extrapolation of our fits down to $x=10^{-6}$ for $Q^2=2,5,10$ and
  $20$ GeV$^2$.}\label{pseudo}
\end{figure}

The analyses of available data, including this work, offer compelling
evidence for the presence of gluon saturation effects. However, a
clear distinction from alternative (linear) approaches cannot yet be
established unambiguously. Indeed, a good description of all data for
$Q^2\geq 1$ can be achieved using DGLAP fits, even though the very
application of DGLAP at such low $Q^2$ is questionable. On the other
hand, the observables like $F_L$ which are more sensitive to the gluon
distribution are poorly determined experimentally. The proposed LHeC
\cite{lhec2} may have the potential to disentangle different physical
scenarios through a deeper exploration of the small-$x$ kinematic
region. The question rises of which observable is best suited to pin
down the saturation of the proton gluon distribution. A partial answer
is given in Fig. \ref{pseudo}, courtesy of Juan Rojo \cite{rojo}. It
presents a NNPDF1.2 DGLAP fit \cite{Ball:2008by,Ball:2009mk} to
pseudodata for proton F$_2$ and F$_L$ in the kinematic regime relevant
for the LHeC (down to $x=10^{-6}$ approx.). The pseudodata are
generated by extrapolation of the fit to F$_2$ presented here using BK
evolution, and thus include saturation effects. While $F_2$ can be
well fitted by DGLAP, the best DGLAP fit to pseudotata underestimate
$F_L$ at small $Q^2$ and overshoots it at the largest $Q^2$
considered, yielding a large $\chi^2/$d.o.f.. One concludes that a
precise experimental determination of $F_L$ at the LHeC over a large
enough $Q^2$ range might suffice to pin down the kinematic region
where departure between DGLAP and non-linear (or, maybe, linear
resummed small-$x$ evolution) takes place, while $F_2$ does not offer
such discrimination power.

\section*{Acknowledgments}
We would like to thank Juan Rojo for providing us with Fig.
\ref{pseudo} and for useful discussions. 

\begin{footnotesize}




\begin{thebibliography}{99}


\bibitem{Kovchegov:2006vj}
Y.~Kovchegov and H.~Weigert, 
 {\em Nucl. Phys. {\bf A}} {\bf 784} (2007) 188--226, {{\tt hep-ph/0609090}}.


\bibitem{Balitsky:2006wa}
I.~I. Balitsky, 
  {\em Phys. Rev. D} {\bf 75} (2007) 014001, {{\tt hep-ph/0609105}}.

\bibitem{Gardi:2006rp}
E.~Gardi, J.~Kuokkanen, K.~Rummukainen, and H.~Weigert, 
{\em Nucl. Phys.} {\bf A784} (2007) 282--340, {{\tt hep-ph/0609087}}.


\bibitem{Albacete:2007yr}
J.~L. Albacete and Y.~V. Kovchegov, 
{\em Phys. Rev.} {\bf
  D75} (2007) 125021, {{\tt arXiv:0704.0612}} [hep-ph].

\bibitem{Golec-Biernat:1998js}
K.~Golec-Biernat and M.~{W{\"u}sthoff}, 
  {\em Phys. Rev.} {\bf D59} (1999) 014017, {{\tt hep-ph/9807513}}.

\bibitem{Albacete:2004gw}
  J.~L.~Albacete, N.~Armesto, J.~G.~Milhano, C.~A.~Salgado and U.~A.~Wiedemann,
  Phys.\ Rev.\  D {\bf 71} (2005) 014003
  {{\tt arXiv:0408216}} [hep-ph].


\bibitem{Albacete:2009fh}
  J.~L.~Albacete, N.~Armesto, J.~G.~Milhano and C.~A.~Salgado,
  {{\tt arXiv:0902.1112}}.

\bibitem{lhec2} M.~Klein {\em et.~al.}, {\it {Prospects for a Large
      Hadron Electron Collider (LHeC) at the LHC}}, EPAC'08, 11th
  European Particle Accelerator Conference, 23- 27 June 2008, Genoa,
  Italy.

\bibitem{Ball:2008by}
  R.~D.~Ball {\it et al.}  [NNPDF Collaboration],
  Nucl.\ Phys.\ B {\bf 809} (2009) 1 [Erratum-ibid.\ B {\bf 816}
  (2009) 293], {{\tt arXiv:0808.1231}} [hep-ph].

\bibitem{Ball:2009mk}
  R.~D.~Ball {\it et al.}  [The NNPDF Collaboration],
  {{\tt arXiv:0906.1958}} [hep-ph].


\bibitem{rojo} J.~Rojo and F.~Caola, arXiv:0906.2079 [hep-ph]. 

\end{thebibliography}
%

\end{footnotesize}


\end{document}